\documentclass[prl,twocolumn]{revtex4}
\usepackage{graphicx}
\usepackage{graphics}
\usepackage{amsfonts}
\usepackage{amsmath}
\usepackage{epsfig}
\usepackage{color}

\begin{document}
\catcode`\ä = \active \catcode`\ö = \active \catcode`\ü = \active
\catcode`\Ä = \active \catcode`\Ö = \active \catcode`\Ü = \active
\catcode`\ß = \active \catcode`\é = \active \catcode`\è = \active
\catcode`\ë = \active \catcode`\ô = \active \catcode`\ê = \active
\catcode`\ø = \active \catcode`\ò = \active \catcode`\í = \active
\catcode`\Ó = \active \catcode`\ú = \active \catcode`\á = \active
\catcode`\ã = \active
\defä{\"a} \defö{\"o} \defü{\"u} \defÄ{\"A} \defÖ{\"O} \defÜ{\"U} \defß{\ss} \defé{\'{e}}
\defè{\`{e}} \defë{\"{e}} \defô{\^{o}} \defê{\^{e}} \defø{\o} \defò{\`{o}} \defí{\'{i}}
\defÓ{\'{O}} \defú{\'{u}} \defá{\'{a}} \defã{\~{a}}



\newcommand{\li}{$^6$Li $ $}
\newcommand{\na}{$^{23}$Na $ $}
\newcommand{\cs}{$^{133}$Cs}
\newcommand{\kk}{$^{40}$K}
\newcommand{\rb}{$^{87}$Rb}
\newcommand{\vect}[1]{\mathbf #1}
\newcommand{\mf}{$m_F$}
\newcommand{\g}{g^{(2)}}
\newcommand{\one}{$|1\rangle$~}
\newcommand{\two}{$|2\rangle$~}
\newcommand{\limol}{$^6$Li$_2$}
\newcommand{\V}{V_{12}}
\newcommand{\kfa}{\frac{1}{k_F a}}
\newcommand{\mixone}{$|1\rangle-|2\rangle$~}
\newcommand{\mixthree}{$|1\rangle-|3\rangle$~}
\newcommand{\mixtwo}{$|2\rangle-|3\rangle$~}
\newcommand{\three}{$|3\rangle$~}

\title{Determination of the fermion pair size in a resonantly interacting superfluid}

\author{Christian H. Schunck, Yong-il Shin, Andr\'{e} Schirotzek, and Wolfgang Ketterle}

\affiliation{Department of Physics\mbox{,} MIT-Harvard Center for
Ultracold Atoms\mbox{,}
and Research Laboratory of Electronics,\\
MIT, Cambridge, MA 02139}

\date{February 4, 2008}

\begin{abstract}
Fermionic superfluidity requires the formation of pairs. The actual
size of these fermion pairs varies by orders of magnitude from the
femtometer scale in neutron stars and nuclei to the micrometer range
in conventional superconductors. Many properties of the superfluid
depend on the pair size relative to the interparticle spacing. This
is expressed in BCS-BEC crossover
theories~\cite{eagl69,legg80,nozi85}, describing the crossover from
a Bardeen-Cooper-Schrieffer (BCS) type superfluid of loosely bound
and large Cooper pairs to Bose-Einstein condensation (BEC) of
tightly bound molecules. Such a crossover superfluid has been
realized in ultracold atomic gases where high temperature
superfluidity has been observed~\cite{zwie05vortex,kett08var}. The
microscopic properties of the fermion pairs can be probed with
radio-frequency (rf) spectroscopy. Previous
work~\cite{chin04,schu07rf,shin07rf} was difficult to interpret due
to strong and not well understood final state interactions. Here we
realize a new superfluid spin mixture where such interactions have
negligible influence and present fermion-pair dissociation spectra
that reveal the underlying pairing correlations. This allows us to
determine the spectroscopic pair size in the resonantly interacting
gas to be $2.6(2)/k_F$ ($k_F$ is the Fermi wave number). The
fermions pairs are therefore smaller than the interparticle spacing
and the smallest pairs observed in fermionic superfluids. This
finding highlights the importance of small fermion pairs for
superfluidity at high critical temperatures~\cite{pist94}. We have
also identified transitions from fermion pairs into bound molecular
states and into many-body bound states in the case of strong final
state interactions.
\end{abstract}

\maketitle

The properties of pairs are revealed in a dissociation spectrum,
where pair dissociation is monitored as a function of the applied
energy $E$. The spectrum has a sharp onset at the pair's binding
energy $E_{b}$, where the fragments have zero kinetic energy, and
then spreads out to higher energy. Since a rf photon has negligible
momentum, the allowed momenta for the fragments reflect the Fourier
transform $\Phi(k)$ of the pair wavefunction $\phi(r)$, which has a
width on the order of $1/\xi$ where $\xi$ is the pair size. Thus the
pair size can be estimated from the spectral line width $E_w$ as
$\xi^2 \sim \hbar^2 / m E_w$ ($m$ is the mass of the particles and
$\hbar$ is Planck's constant $h$ divided by $2\pi$).

\begin{figure}
\begin{center}
\includegraphics[width=3.2in]{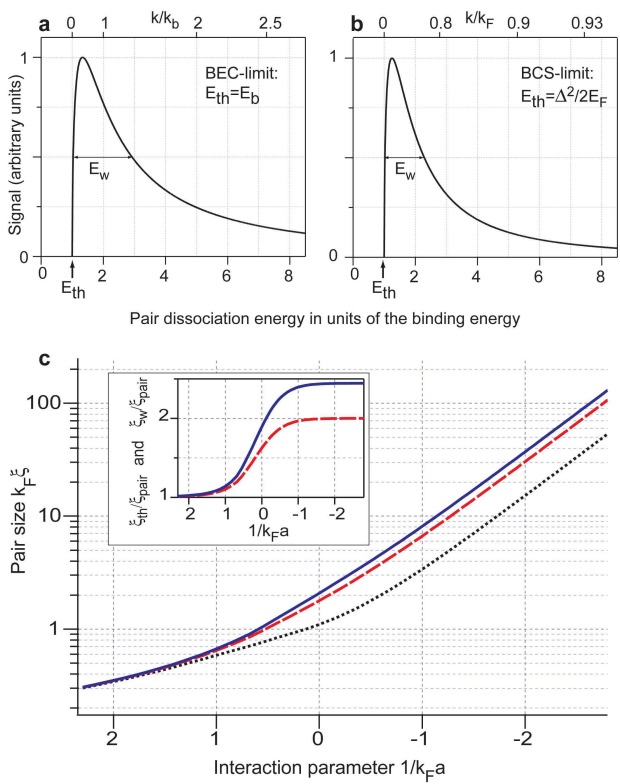}
\caption[Title]{Line shape of the pair dissociation spectrum in the
BEC (\textbf{a}) and BCS limit (\textbf{b}) and the evolution of the
fermion pair size in the BEC-BCS
crossover~\cite{enge97,dien04rf,kett08var}. (\textbf{a}) and
(\textbf{b}): Simulated rf dissociation spectra in the BEC and BCS
limits. The momentum $k$ of the free particles after dissociation is
indicated in the top axes, where $\hbar^2 k_b^2 /m=E_b$. Apart from
an offset, the spectra in the BEC and BCS limits show almost
indistinguishable lineshapes. The molecular dissociation lineshape
$I_m$ with an additional offset parameter can therefore serve as a
generic, model independent fit function for pair dissociation
spectra (see Methods and Fig.~\ref{fig:sfig3}). (\textbf{c}) The
fermion pair sizes $\xi_{w}$ (solid blue) and $\xi_{th}$ (dashed
red) are displayed as a function of the interaction parameter
$1/k_Fa$ ($a$ is the $s$-wave scattering length). Also shown is the
two-particle correlation length $\xi_{pair}$ (dotted black) given by
$\xi_{pair}=\sqrt{\langle\phi|r^2|\phi\rangle/\langle\phi|\phi\rangle}$,
where $\phi(r)=\langle \psi| \Psi_{\alpha}^\dag (r)
\Psi_{\beta}^\dag (0) |\psi\rangle$. Here $\psi$ is the generalized
BCS wavefunction and $\alpha$ and $\beta$ refer to the two
components~\cite{legg80}. In the BEC limit, the value for the
molecular size is $\xi_m = b/\sqrt{2}=\xi_{pair}$. We chose
$\gamma=1.89$ in the definition of $\xi_w$, so that
$\xi_m=\xi_{th}=\xi_w$. In the BCS limit,
$\xi_{pair}=\pi/(2\sqrt{2})\xi_c$ where $\xi_c=\hbar^2 k_F/(\pi m
\Delta)$ is the Pippard coherence length and we have
$\xi_{th}=2\xi_{pair}$, $\xi_w=2.44 \xi_{pair}$. The inset shows the
ratios $\xi_{w}/\xi_{pair}$ (solid blue) and $\xi_{th}/\xi_{pair}$
(dashed red). Although $\xi_{pair}$ changes by orders of magnitude,
$\xi_{th}$ and $\xi_w$ show the same behavior as $\xi_{pair}$
deviating from each other by not more than 22 \%. This illustrates
that the pair size can be reliably determined from the rf
dissociation spectrum throughout the whole BEC-BCS crossover.}
\label{fig:fig1}
\end{center}
\end{figure}

\begin{figure*}
\begin{center}
\includegraphics[width=3in]{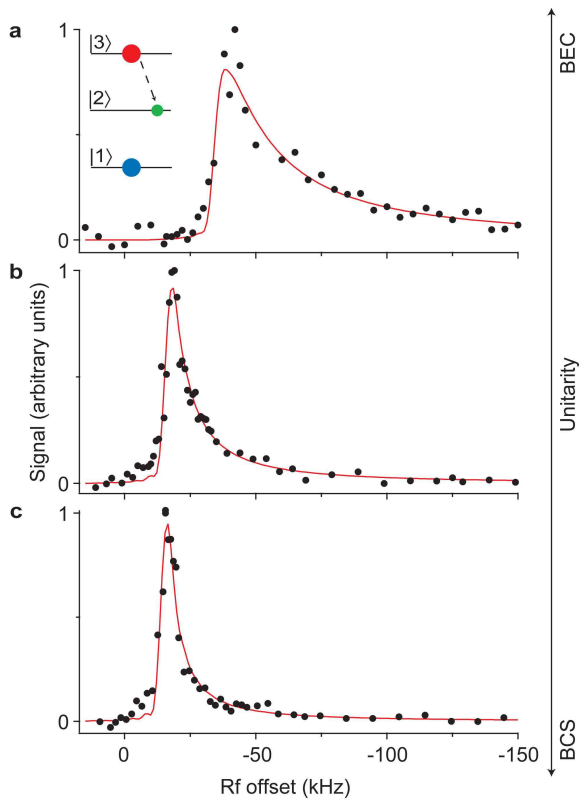}
\caption[Title]{Rf dissociation spectra in the BEC-BCS crossover.
Below, at, and above resonance, the spectrum shows the typical
asymmetric lineshape of a pair dissociation spectrum. The signal is
proportional to the three dimensional local response at the center
of the cloud (see Methods). Since state $|3\rangle$ has a higher
energy than state $|2\rangle$ (see the schematic inset in
(\textbf{a})), the dissociation energy is always less than the
transition frequency for the atomic resonance $E_0/h$ and therefore
the dissociation spectra appear at negative energies compared to
$E_0$. The inverted frequency axis ensures that the dissociation
spectrum is always on the right (or ``positive'') side of the
origin. The magnetic field (in G), the local Fermi energy
$\epsilon_F$ (in kHz), the temperature $T$ in units of the Fermi
temperature $T/T_F$ and the interaction strength $1/k_Fa_i$ are
(\textbf{a}), 670, $h \times 24$, $\approx 0.2$, 0.4; (\textbf{b}),
691, $h \times 21$, 0.1, $\sim0$; (\textbf{c}), 710, $h \times 20$,
0.1, -0.3.} \label{fig:fig2}
\end{center}
\end{figure*}

\begin{figure}
\begin{center}
\includegraphics[width=2.6in]{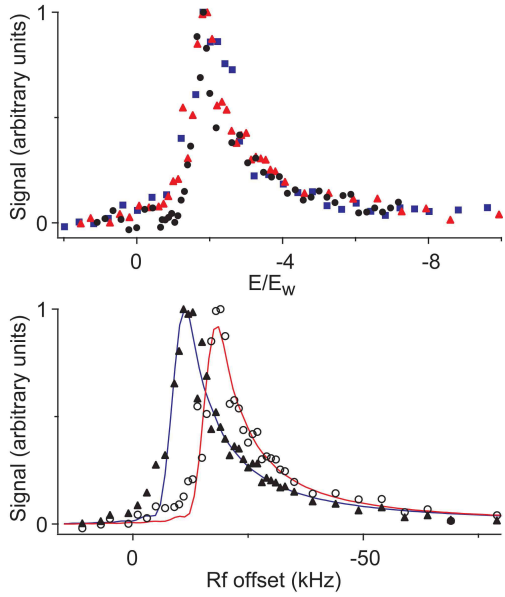} \caption[Title]{Comparison of
lineshapes and density effects. (\textbf{a}) Same spectra as in
Fig.~\ref{fig:fig2} but with the frequency axis scaled by $E_{w}$
and shifted so that the spectral onsets overlap with the BEC side
spectrum: BEC side (black circles), resonance (red triangles), and
BCS side (blue squares). (\textbf{b}) Density effects at unitarity
for the (1,3) mixture at 691 G. The figure shows the tomographically
reconstructed spectral response in the center (open circles, same
spectrum as in Fig.~\ref{fig:fig2}b) as well as the lower density
wings (filled triangles) of the cloud. In this regime the cloud
might have turned normal.} \label{fig:fig3}
\end{center}
\end{figure}

\begin{figure}
\begin{center}
\includegraphics[width=2.6in]{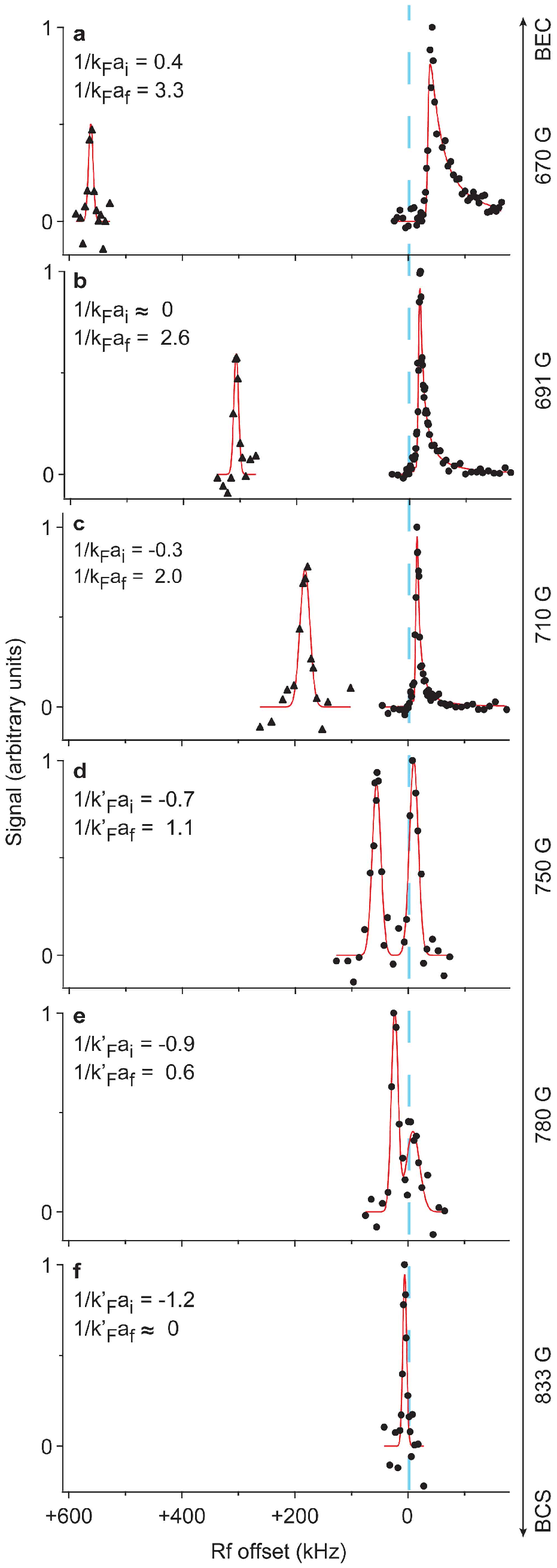}
\caption[Title]{Effect of final state interactions on rf
spectroscopy: bound-bound (BB) and bound-free (BF) spectra in the
BEC-BCS crossover of the (1,3) mixture (only the BF spectra in
(\textbf{a}-\textbf{c}) were tomographically reconstructed). While
the initial (1,3) state is strongly interacting at all fields the
final state interactions change from weak (\textbf{a}-\textbf{c}) to
strong (\textbf{d}-\textbf{f}). See ref.~\cite{bart04fesh} for a
plot of the Feshbach resonances. At the higher magnetic fields for
$1/k_Fa_i\approx-1$ the initial state may have turned normal.
(\textbf{a}-\textbf{c}), Same BF spectra and parameters as in
Fig.~\ref{fig:fig2}. The relative weight of the BB and BF peaks
could not be determined experimentally (see Methods).(\textbf{d}),
750 G, $E_F=h \times 22$ kHz, $T/T_F$=0.09; (\textbf{e}), 780 G,
$E_F=h \times 23$ kHz, $T/T_F$=0.09; (\textbf{f}), 833 G, $E_F=h
\times 20$ kHz, $T/T_F$=0.06.} \label{fig:fig4}
\end{center}
\end{figure}

The conceptually simplest pairs in the BCS-BEC crossover are the
weakly bound molecules in the BEC limit, which are described by a
spatial wavefunction $\phi_{m}(r) \propto e^{-r/b}/r$ with a binding
energy $E_b=\hbar^2/ m b^2$. When the molecules are dissociated into
non-interacting free particles, the spectral response is $I_m
\propto \sqrt{E-E_b}/E^2$, showing a highly asymmetric line shape
with a steep rise at the molecular binding energy $E_{b}$ and a long
``tail'' to higher energies
(Fig.~\ref{fig:fig1}a)~\cite{kett08var,rega03mol}.

This general behavior of the dissociation spectrum holds also in the
BCS limit where pairing is a many-body
effect~\cite{dien04rf,kett08var}. The rf dissociation process
discussed below, in the limit of negligible final state
interactions, can be considered as breaking a Cooper pair into one
quasiparticle and one free particle. The rf spectrum in the BCS
limit has an onset at $\Delta^2/2E_F$ and the same dependence of
$E^{-3/2}$ at high energy as in the BEC limit (Fig.~\ref{fig:fig1}b;
here $E_F$ is the Fermi energy and $\Delta$ is the
gap)~\cite{yu06delta}. Since the rf excitation takes place
throughout the whole Fermi sea it is most natural to interpret the
BCS state as $N/2$ pairs with condensation energy $\Delta^2/2E_F$
where $N$ is the total number of fermions~\cite{kett08var}.

A spectroscopic pair size can be defined both from the onset and the
width of the rf spectrum as $\xi_{th}^2 = \hbar^2/2 m E_{th}$ and
$\xi_w^2 = \gamma \times \hbar^2/2 m E_w$. Here $E_{th}$ is the
onset/threshold energy, $E_{w}$ is the full width at half maximum,
and $\gamma=1.89$ is a numerical constant chosen for convenience
(see caption of Fig.~\ref{fig:fig1}). The pair sizes $\xi_{th}$ and
$\xi_w$ which can be directly obtained from the rf spectrum capture
the evolution of the pair size from the BCS limit to the BEC limit
(see Fig.~\ref{fig:fig1}c).

Since the rf spectra show a similar behavior in both limiting cases
of the BEC-BCS crossover, one would expect comparable spectra within
the crossover regime. Surprisingly, the rf spectra obtained in
previous rf experiments did not fit into this picture: the lineshape
did not show any pronounced asymmetry and the linewidth was
narrow~\cite{chin04,schu07rf,shin07rf} (see also
Fig.~\ref{fig:sfig4}). These experiments could therefore not be
simply interpreted in terms of pairing energy and pair size. We will
show that this is caused by strong final state interactions and
transitions to bound states.

In the previous and our new experiments, the fermion pairs consist
of two atoms in different hyperfine states $|a\rangle$ and
$|b\rangle$. The rf transfers atoms in state $|b\rangle$ to an
initially unoccupied third state $|c\rangle$. In addition to ``pair
dissociation'', also referred to as a ``bound-free'' transition and
characterized by the asymmetric lineshape discussed above, rf
spectroscopy can induce a second kind of transition to another bound
state, i.e. the transfer of a pair (a,b) to a pair (a,c)  (also
referred to as a ``bound-bound'' transition). The latter spectra
have a narrow and symmetric lineshape.

Final state effects arise when the dissociated atom in state
$|c\rangle$ interacts with atoms in state $|a\rangle$. The
interaction strength is measured by the dimensionless parameter
$k_Fa$. Here $a$ is the $s$-wave scattering length and we use $a_i$
($a_f$) for the initial (a,b) (final (a,c)) interactions. As
discussed in detail below, final state interactions severely affect
the rf dissociation spectra when $|k_Fa_f|>1$
~\cite{baym07rf,punk07rf,pera08rf}. To overcome this problem, one
has to change the interactions in the final state without changing
those in the initial one. Our solution is the realization of a new
high temperature superfluid in \li using a different combination of
hyperfine states for which rf excitation with reduced final state
interactions is possible (see Methods). As a result, we were able to
resolve the bound-bound and bound-free contributions to the rf
spectrum, and to determine the size of fermion pairs from the
asymmetric fermion pair dissociation spectra.

We have taken advantage of the fact that any two state mixture
(1,2), (1,3), and (2,3) of the three lowest hyperfine states of \li
(labeled in the order of increasing hyperfine energy as $|1\rangle$,
\two and $|3\rangle$) exhibits a broad Feshbach
resonance~\cite{gupt03rf,bart04fesh}. So far, all experiments with
strongly interacting fermions in \li have been carried out in the
vicinity of the (1,2) Feshbach resonance located at about $B_{12}
\sim 834$ G. Surprisingly, inelastic collisions including allowed
dipolar relaxation are not enhanced by the (1,3) and (2,3) Feshbach
resonances. We observe that at both the (1,3) and (2,3) Feshbach
resonances superfluids can be created as well (see Methods). This
doubles the number of high temperature superfluids available for
experimental studies.

The newly created (1,3) superfluid is the best choice for rf
spectroscopy experiments since the final state scattering length
$a_f$ at the (1,3) resonance position $B_{13} \sim 691$ G is small
and positive ($0<k_Fa_f<1$ for typical values of $k_F$). Therefore
the accessible final states are either a molecule of a well defined
binding energy or two free, only weakly interacting atoms. The
actual final state interactions depend on whether one drives the rf
transitions from \one to \two or from \three to $|2\rangle$ allowing
the comparison between spectra taken from the same sample but with
different $a_f$ (see Methods and Supplementary Information). After
preparing the (1,3) superfluid a rf pulse resonant with the \three
to \two transition is applied. Then either the losses in state
\three or the atoms transferred to state \two are monitored (see
Methods). All spectra are plotted versus frequency or energy
relative to the atomic resonance, i.e.~relative to the energy $E_0$
required to transfer an atom from $|3\rangle$ to $|2\rangle$ in the
\textit{absence} of atoms in state $|1\rangle$.

The main result of this paper are the spectra observed in the (1,3)
BEC-BCS crossover between 670 and 710 G (Fig.~\ref{fig:fig2}). The
spectra have the asymmetric lineshape characteristic for pair
dissociation and are indeed well fit by a generic pair dissociation
lineshape convolved with the lineshape of the square excitation
pulse (see Fig.~\ref{fig:fig1} and Methods). If the frequency axis
is scaled by $E_{w}$ and the spectra are shifted to show the same
onset all three spectra overlap as shown in Fig.~\ref{fig:fig3}a. At
the level of our experimental resolution the dissociation lineshape
is therefore not sensitive to a change in interactions. As
illustrated in Fig.~\ref{fig:fig1}, the pair size can in principle
be obtained from both $E_{th}$ and $E_{w}$. However, since the whole
spectrum may be subject to shifts from Hartree
terms~\cite{rega03,gupt03rf}, we focus in the following only on the
width of the spectrum.

At unitarity we determine the full width at half maximum to be
$E_w=0.28(5) \epsilon_F$ corresponding to a spectroscopic pair size
of $\xi_w=2.6(2)/k_F$ (here $\epsilon_F$ is the local Fermi energy
and $k_F=\sqrt{2m\epsilon_F}/\hbar$; the quoted errors are purely
statistical). The pairs are therefore smaller than the interparticle
spacing $l$ given by $l = n^{1/3}=(3\pi^2)^{1/3}/k_F \sim 3.1/k_F$
(where $n$ the total density) and in units of $1/k_F$ the smallest
reported so far for fermionic superfluids. In high-temperature
superconductors the reported values for $\xi$ at optimal doping are
in the range of 5 to $10/k_F$~\cite{pist94}.

In the simple BEC-BCS crossover model the ratio $\xi_{pair}/\xi_w$
varies from 1 to $1/2.4$. The fact that $\xi_w$ is smaller than $l$
suggests the use of the molecular ratio, i.e.
$\xi\equiv\xi_{pair}=\xi_w=2.6/k_F$. Before we compare with
theoretical predictions we note that various definitions of the pair
size differ by factors on the order of unity~\cite{orti05pair}. With
this in mind, we find that our observed $\xi$ is larger than a
predicted pair size of about $1/k_F$ based on a functional integral
formulation of the BEC-BCS crossover~\cite{enge97}. Small fermion
pair sizes have been explicitly linked to high critical temperatures
via the relation $T_c/T_F \approx 0.4/(k_F\xi_{pair})$ which applies
for weak coupling~\cite{pist94}. Inserting the observed $\xi$ this
relation yields an estimate of $T_c/T_F\approx0.15$ which is in the
range of the predicted values between 0.15 to 0.23 (here $T_F$ is
the local Fermi temperature)~\cite{buro06}. If we use the asymptotic
BCS relation $\Delta=\frac{\hbar^2k_F}{\pi
m}\frac{1}{\xi_c}=\frac{1}{\sqrt{2}}\frac{1}{k_F\xi_{pair}}
\epsilon_F$, valid at weak coupling, and our observed $\xi$ at
unitarity we find $\Delta\approx0.3 \epsilon_F$ smaller than the
value of $0.5\epsilon_F$ predicted by Monte Carlo
simulations~\cite{carl06}.

The strong narrowing of the spectral line in Fig.~\ref{fig:fig2} (a)
to (c) demonstrates that the fermion pair size increases from strong
to weak coupling. The decreasing width corresponds to a more than
twofold increase of the spectroscopic pair size from
$\xi_w=1.4(1)/k_F$ at 670 G to $\xi_w=3.6(3)/k_F$ at 710 G. A change
of the absolute pair size with density at unitarity can in principle
be observed by comparing the spectral width in the center and the
outer region of the cloud. As the density decreases, the spectrum
shifts to lower energies (see Fig.~\ref{fig:fig3}b). However, the
spectral onset also becomes increasingly softer and the asymmetry of
the pair dissociation peak less pronounced, possibly due to atomic
diffusion during the excitation pulse. This prevents a reliable
determination of the pair size in the spatial wings where the
density is changing rapidly.

We now consider the effect of final state interactions in more
detail. First we would like to point out that the increase in $a_f$
by about a factor of two from 670 G to 710 G has not affected the
lineshape of the spectra in Fig.~\ref{fig:fig3}a within the
experimental resolution. This suggests that final state effects are
small for these spectra. Additional information is obtained from the
previously introduced bound-bound (BB) transitions which are outside
the range plotted in Fig.~\ref{fig:fig2}. On the BEC side of the
resonance the (1,3) molecule can be transferred also to a more
deeply bound (1,2) molecule (see Fig.~\ref{fig:fig4}a). The BB peak
is still present at unitarity and also on the BCS side at 710 G
(Fig.~\ref{fig:fig4}b and c) and results from the transition of a
many-body bound fermion pair to a (1,2) molecule. The strong overlap
of the pair wavefunction and the molecule in the final state is
another indication for the ``molecular'' character of the fermions
pairs in the strongly interacting regime.

The spectra start to change significantly at higher fields. As the
magnetic field is increased the (1,3) mixture remains in the
unitarity limited regime with the interaction strength approaching
$1/|k_Fa_i|\approx1$ at $B_{12}=833$ G. The final state
interactions, however, change from weak to strong causing the pair
dissociation peak to decrease in weight and the BB peak to become
dominant (Fig.~\ref{fig:fig4}d-f). This single peak apparently
corresponds to a BB transition from many-body bound (1,3) pairs to a
highly correlated final state of an atom in state $|2\rangle$
interacting with the paired atoms in state $|1\rangle$.

A narrow BB peak is predicted both in the molecular (two-body) and
many-body case, when initial and final state interactions are
identical or similar. The spectra in Fig.~\ref{fig:fig4} show that
BB transitions dominate when
$\left|1/(k_Fa_i)-1/(k_Fa_f)\right|\leq1.5$. In our opinion a recent
theoretical treatment~\cite{basu07rf} agrees qualitatively with
these results but underestimates the region where BB transitions are
dominant by about a factor of two. Our observations allow a
reinterpretation of the rf spectra obtained from the (1,2)
superfluid with resonant
interactions~\cite{chin04,schu07rf,shin07rf} (see the Supplementary
Information for an extended discussion). The spectra have been taken
in a regime where $\left|1/(k_Fa_i)-1/(k_Fa_f)\right|\leq1$ where
strong BB transitions are expected. Together with the very narrow
and symmetric lineshape (see Fig.~\ref{fig:sfig4}), this suggests
that the (1,2) to (1,3) rf spectra at 833 G are dominated by such BB
transitions and cannot be simply interpreted in terms of a pair
dissociation process and a pairing
gap~\cite{chin04,kinn04gap,ohas05,he05rf,shin07rf,schu07rf}.

In conclusion we have determined the pair size of resonantly
interacting fermions using new superfluid spin mixtures in $^6$Li.
The (1,3) mixture is ideally suited for rf spectroscopy since final
state interactions do not significantly affect the spectra. Our
measurements are the first to clearly reveal the microscopic
structure of the fermion pairs in the strongly interacting regime.
The small fermion pair size and high critical temperatures observed
in our system show a relation similar to the one suggested by the
Uemura plot for a wide class of fermionic superfluids~\cite{pist94}.
Our results also explain why the rapid ramp method used to observe
fermion pair condensation in the crossover has been
successful~\cite{rega04,zwie04pairs}. The small pair size
facilitated the efficient transfer of the many-body bound fermion
pairs into more strongly bound molecules while preserving the
momentum distribution of the pairs.

This work opens ample opportunities for future research. The
microscopic structure of the pairs can now be studied both in the
superfluid and normal phase as a function of interaction strength,
temperature and spin imbalance between the two
components~\cite{schu07rf}. Increased spectral resolution may reveal
interesting deviations of the spectral shape from the generic
lineshape discussed here. Furthermore, the predicted universality of
a resonantly interacting Fermi mixture can now be tested in \li for
three different systems. The lifetimes of all three two-state
combinations of the three lowest hyperfine states in \li are on the
order of 10 s in the strongly interacting regime. The three-body
decay rates, however, decrease by more than an order of magnitude
between 690 and 830 G for a ternary mixture, which may reflect
interesting three-body physics. The lifetime of 30 ms at 691 G might
be sufficient for studies involving all three hyperfine
states~\cite{hone04} with the potential for experiments on pairing
competition in multi-component Fermi gases and spinor Fermi
superfluids.

We thank M. Zwierlein, W. Zwerger, E. Mueller and S. Basu for
stimulating discussions and A. Keshet for the experiment control
software. This work was supported by the NSF and ONR, through a MURI
program, and under ARO Award W911NF-07-1-0493 with funds from the
DARPA OLE program.

\vspace{0.3in} \noindent \textbf{Methods Summary}

\begin{figure}
\begin{center}
\includegraphics[width=2.6in]{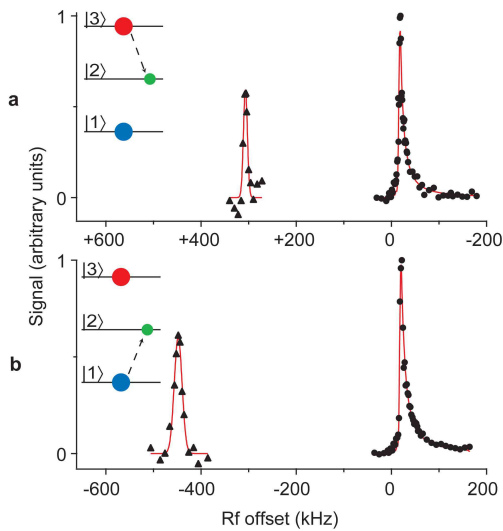} \caption[Title]{
Rf spectra at unitarity for the (1,3) mixture at 691 G. For all
spectra the number of atoms transferred to state $|2\rangle$ has
been recorded. In the (1,3) mixture rf transitions to the final
state $|2\rangle$ are possible from both states $|1\rangle$ and
$|3\rangle$. The final states can therefore be either bound (2,3) or
(1,2) molecules respectively or a dissociated free atom in state
$|2\rangle$. Note that the bound-free (BF) spectra are very similar
for both $|3\rangle$ to $|2\rangle$ (\textbf{a}) and $|1\rangle$ to
$|2\rangle$ (\textbf{b}) transitions. The bound-bound (BB) spectra,
however, show different shifts indicating that the final (2,3)
molecule is more strongly bound than the (1,2) molecule. This is a
consequence of the smaller width of the (2,3) Feshbach resonance at
811 G~\cite{bart04fesh} compared to the width of the (1,2) resonance
at 834 G. (\textbf{a}) \three to \two transition; $\epsilon_F=21$
kHz, $T/T_F$=0.1 (\textbf{b}) \one to \two transition;
$\epsilon_F=22$ kHz, $T/T_F$=0.14.} \label{fig:sfig2}
\end{center}
\end{figure}

\textbf{Creation of the (1,3) superfluid.} As described
previously~\cite{kett08var} a spin polarized sample of ultracold \li
in state \one is obtained in an an optical dipole trap after
sympathetic cooling with \na in a magnetic trap. The equal (1,3)
mixture is prepared at 568 G, close to the zero crossing of
$a_{13}$. Here a non-adiabatic Landau-Zener rf sweep, creating an
equal (1,2) mixture, is followed by an adiabatic Landau-Zener sweep
that transfers the atoms in state \two to state \three. To induce
strong interactions the magnetic field is adjusted in 100 ms to 730
G and then ramped to values between 660 and 833 G. After evaporative
cooling in the optical trap, superfluidity is indirectly established
via the observation of fermion pair
condensates~\cite{rega04,zwie04pairs}. Under comparable conditions
quantized vortex lattices, a direct proof for superfluidity, have
been observed in the rotating (1,2) mixture of
$^6$Li~\cite{zwie05vortex}.  $E_F =
h(\nu_r^2\nu_{ax})^{1/3}(3N)^{1/3}$ with radial (axial) trapping
frequencies $\nu_r=140$ ($\nu_{ax}=22$) Hz and
$k'_F=\sqrt{2mE_F}/\hbar$. The temperature was determined from the
shape of the expanded cloud.

\textbf{Recording the (1,3) rf spectra.} The rf dissociation spectra
at 670, 691 and 710 G spectra have been obtained by applying a 200
$\mu$s long rf pulse to the (1,3) mixture monitoring the atoms
transferred into state $|2\rangle$. Three-dimensional image
reconstruction via the inverse Abel transformation was used to
obtain local rf dissociation spectra~\cite{shin07rf}. The pulse
length was chosen to be shorter than 1/4 trapping period to minimize
atomic diffusion during the excitation pulse. The rf power was
adjusted to transfer less than 5\% of the total number of atoms. A
further reduction of the rf power only affected the signal to noise
ratio but not the spectral width. All BB spectra and the spectra at
fields at and above 750 G are not spatially resolved and were
obtained with about 1 ms long rf pulses.

\vspace{0.1in} \noindent \textbf{Full methods}

\textbf{Creation of new superfluid spin mixtures for rf
spectroscopy.} For the well established (1,2) mixture, only the
$|2\rangle$ to $|3\rangle$ transition has been used for rf
spectroscopy. The final state $s$-wave scattering length $a_{13}$ at
$B_{12}$ is large and negative leading to strong final state
interactions with $1/k_Fa_f<-1$ ($a_{13}\approx-3300$ $a_0$, $a_0$
the Bohr radius and $a_{ij}$ the magnetic field dependent scattering
length between atoms in states $|i\rangle$ and $|j\rangle$). The
strength of the final state interactions can in principle be changed
in several ways without affecting the initial state. The density
could be lowered to reduce the interaction strength in the final
state while the initial state remains resonantly interacting. It is,
however, experimentally difficult to decrease the density by a large
factor and maintain the same low temperature $T/T_F$. One might also
try to spectroscopically access a different final state. However, in
\li there are no other allowed magnetic field insensitive
transitions. Magnetic field insensitivity is necessary to obtain the
required spectral resolution in the kHz regime.

Since other mixtures of hyperfine states in \li also exhibit broad
Feshbach resonances we attempted to create resonantly interacting
superfluids in new combinations of initial hyperfine states: (1,3)
and (2,3). The lifetimes of these spin mixtures at resonance exceed
10 s implying inelastic collision rates smaller than $10^{-14}$
cm$^{-3}$ s$^{-1}$. While for the (2,3) superfluid the final state
interactions are also large and negative, the final state scattering
lengths at $B_{13}$ are either $a_{23}\approx 1140$ $a_0$ and
$a_{12}\approx 1450$ $a_0$ (depending on the rf transition employed)
and therefore considerably smaller and positive.

\textbf{Creation of the (2,3) superfluid.} To prepare a (2,3)
superfluid we follow essentially the same procedure as previously
described for the (1,2) mixture~\cite{zwie05vortex,kett08var}. The
only difference is that instead of applying a Landau-Zener transfer
that creates an equal (1,2) mixture a complete transfer into state
\two is followed by a second sweep creating an equal (2,3) mixture.
The final magnetic field at the center of the (2,3) resonance is
$B_{23}\approx 811$ G. As in the other spin mixtures we observe
fermion pair condensation after evaporation in the optical trap.

\textbf{Recording the (1,3) spectra: stability of the mixture after
the rf pulse} Recording the atoms transferred to state $|2\rangle$
is advantageous because there is no background without rf pulse, but
it requires that their lifetime with respect to three-body
recombination is sufficiently long.

For fields below $\sim710$ G, we found that the lifetime of the
$|2\rangle$ atoms after the rf pulse was short when they formed a
molecule with an $|1\rangle$ atom as the result of a BB transition.
Therefore, in some cases, the BB part of the spectra was recorded by
observing atom number loss in the initial state. After BF
(bound-free) excitation, the lifetime of atoms in state $|2\rangle$
was 30 ms (determined at 691 G) sufficiently long to observe the
atoms directly. As a result of the different decay times and
recording methods the relative signal strength between the BB and BF
parts of the spectrum could not be determined.

At fields above $\sim750$ G, we found similar and strong losses
after both BF and BB excitations. Therefore all data were taken by
monitoring losses in the initial state $|3\rangle$ and the spectra
reflect the relative strength between BB and BF transitions.

\textbf{Fitting the (1,3) spectra} The fit to the (1,3) rf
dissociation spectra in
Figures~\ref{fig:fig2},~\ref{fig:fig3},~\ref{fig:fig4}(a-c),~\ref{fig:sfig2},
and~\ref{fig:sfig4} uses a generic, model independent pair
dissociation lineshape based on $I_m$ with an additional parameter
$E_{\rm{offset}}$: $I_{\rm{generic}}(E)\propto
\sqrt{(E-E_{th}}/(E-E_{\rm{offset}})^2$. We used this lineshape
convolved with the Fourier transform of the square pulses as a
fitting function, and found good agreement with the experimentally
obtained spectra shown in Fig.~\ref{fig:fig2}. The generic fit
function contains no corrections for final state interactions. In
the BEC limit (where $E_{\rm{offset}}=0$ and $E_{th}=E_b$) such
corrections can be included by a multiplicative factor of
$1/(E+\hbar^2/(m a_f^2)-E_{th})$~\cite{chin05rf}. When applied to
the dissociation spectra in the crossover this correction factor
changes the fit only by a negligible amount. All BB and BF spectra
in Fig.~\ref{fig:fig4}(d-f) have been fit by a Gaussian.

\begin{figure}
\begin{center}
\includegraphics[width=2.6in]{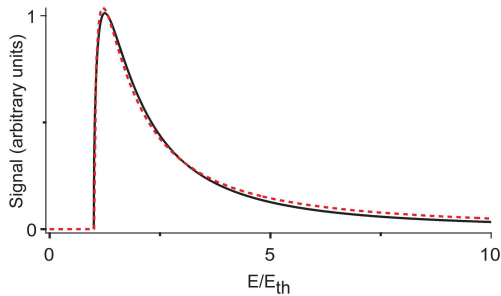} \caption[Title]
{Generic pair dissociation lineshape. A simulated rf dissociation
spectrum in the BCS limit (black solid line)~\cite{kett08var} is fit
with $I_{\rm{generic}}$ (red dashed line) which is the molecular
lineshape $I_m$ with an additional offset parameter
$E_{\rm{offset}}$ (see Methods).} \label{fig:sfig3}
\end{center}
\end{figure}

\textbf{(1,3) mixture: $|3\rangle$ to $|2\rangle$ vs $|1\rangle$ to
$|2\rangle$ transition} The (1,3) superfluid gives us the
opportunity to record two different (magnetic field insensitive) rf
spectra: from state $|3\rangle$ to state $|2\rangle$ (the transition
used for all the spectra shown in the paper) and from state
$|1\rangle$ to state $|2\rangle$. This allows us to compare rf
spectra of the same system but for somewhat different final state
interactions. The final state scattering lengths at $B_{13}$ are
$a_{23}\approx 1140$ $a_0$ for the $|1\rangle$ to $|2\rangle$
transition and $a_{12}\approx 1450$ $a_0$ for the $|3\rangle$ to
$|2\rangle$ transition. Figure~\ref{fig:sfig2} shows the spectra at
691 G. Note that the fermion pair size obtained from the spectra
agrees for both rf transitions within the experimental uncertainty.

\vspace{0.3in} \noindent \textbf{Supplementary Information}

\begin{figure}
\begin{center}
\includegraphics[width=2.6in]{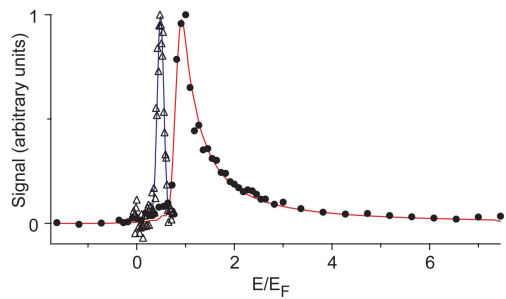} \caption[Title]
{Comparison of the rf spectra of the (1,2) and (1,3) superfluids at
unitarity, showing dramatic final state effects for the (1,2)
mixture. Open circles: same rf dissociation data as in
Fig.~\ref{fig:sfig2}b. Solid diamonds: rf spectra at unitarity for
the (1,2) mixture at 833 G from ref.~\cite{shin07rf}. The frequency
axis is normalized by the local Fermi energies. In the (1,2) mixture
final state effects lead to a strong suppression of the asymmetric
``tails'' of the rf spectrum and a shift of the peak to lower
energies.} \label{fig:sfig4}
\end{center}
\end{figure}


\textbf{Final state interactions in the rf spectroscopy experiments
with the (1,2) and (1,3) mixtures.} Figure~\ref{fig:sfig4} shows the
dramatic effect of final state interactions in the (1,2) mixture at
unitarity. The narrow and symmetric lineshape observed in the (1,2)
to (1,3) rf spectrum suggests that this spectral peak is dominated
by a bound-bound (BB) transition from (1,2) pairs to a (1,3)
correlated state.

In the molecular case final state interactions can be included in an
analytical model~\cite{chin05rf}. The final states for dissociation
are two atoms with momentum $\hbar k$ in an $s$-wave scattering
state with scattering length $a_f$. For a large and positive $a_i
\approx b$ and an increasing $a_f$ ($0 < a_f < a_i \approx b$) the
dissociation spectrum looses in weight and narrows as
$(1-a_f/a_i)/(1+k^2a_f^2)$ until it disappears when $a_f/a_i$
approaches one. At this point the spectrum consists of a delta
function for the BB transition between molecular states of equal
size.

A very similar behavior of the BB and bound-free (BF) parts of the
spectrum is expected for a superfluid with resonant
interactions~\cite{basu07rf}: for $|a_f|, |a_i| \gg 1/k_F$ the
spectrum is reduced to a delta function. Here, the initial state is
a fermion pair condensate described by the BEC-BCS crossover
wavefunction~\cite{basu07rf,kett08var}. In contrast to the molecular
case, the spectrum of the superfluid at resonance shows a BB peak
even for negative values of $1/k_F a_f$, i.e. in a regime where
binding is only due to many-body effects~\cite{basu07rf}. The
spectra in Fig.~\ref{fig:fig4} show that BB transitions dominate
when $\left|1/(k_Fa_i)-1/(k_Fa_f)\right|\leq1.5$ (a region that is
about a factor of two larger than obtained in ref.~\cite{basu07rf}).
We also infer from~\cite{basu07rf} that it is much more difficult to
spectrally resolve BB and BF transitions for a system in the
unitarity limit if $a_f<0$. When one approaches resonance for the
(1,2) system from the BEC side the BF spectrum narrows and smoothly
turns into a BB dominated spectrum.

Compared to the (1,2) superfluid, $a_f$ in the (1,3) system is up to
three times smaller and positive. This leads, both in the molecular
model (due to the quadratic dependence on $k_Fa_f$) and in the
resonant case~\cite{basu07rf}, to a dramatic change in the
dissociation spectrum towards the limit of negligible final state
interactions. Fits to the (1,3) dissociation spectrum both with and
without a correction factor for final state effects (see
Methods)~\cite{chin05rf} show negligible differences, indicating the
small influence of final state interactions. In fact the (1,3)
spectra in Fig.~\ref{fig:fig4}(a-c) show the absence of final state
effects without any detailed analysis. The splitting between BB and
BF parts given by $\hbar^2/m a_f^2$ is considerably larger than the
width of the BF spectrum (which is approximately $\hbar^2/m b^2$).
Therefore the condition $a_f < b$ is fulfilled, implying that final
state interactions do not strongly affect the dissociation spectrum.


\end{document}